\documentstyle[aps,preprint,pra]{revtex}
\begin{document}
\newcommand{\etal}{{\em et al.}\/}
\newcommand{\IP}{inner polarization}
\newcommand{\IPF}{\IP\ function}
\newcommand{\IPFs}{\IP\ functions}
\newcommand{\jcite}[4]{{\it #1} {\bf #4}, {\it #2}, #3}
\newcommand{\auth}[2]{#2, #1;}
\newcommand{\oneauth}[2]{#2, #1;}
\newcommand{\edit}[2]{#2, #1, Ed.}
\newcommand{\twoedit}[4]{#2, #1; #4, #3, Eds.}
\newcommand{\inpress}[1]{{\it #1}, in press}
\newcommand{\subm}[1]{{\it #1}, submitted}
\newcommand{\et}{}
\newcommand{\twoauth}[4]{#2, #1; #4, #3;}
\newcommand{\andauth}[2]{#2, #1;}
\newcommand{\BOOK}[4]{{\it #1}; #2: #3, #4}
\newcommand{\editedbook}[5]{{\it #1}; #2; #3: #4, #5}
\newcommand{\inbook}[5]{In {\it #1}; #2; #3: #4, #5}
\newcommand{\tbp}{to be published}
\newcommand{\erratum}[3]{\jcite{erratum}{#1}{#2}{#3}}
\newcommand{\JCP}[3]{\jcite{J. Chem. Phys.}{#1}{#2}{#3}}
\newcommand{\jms}[3]{\jcite{J. Mol. Spectrosc.}{#1}{#2}{#3}}
\newcommand{\jmsp}[3]{\jcite{J. Mol. Spectrosc.}{#1}{#2}{#3}}
\newcommand{\theochem}[3]{\jcite{J. Mol. Struct. ({\sc theochem})}{#1}{#2}{#3}}
\newcommand{\jmstr}[3]{\jcite{J. Mol. Struct.}{#1}{#2}{#3}}
\newcommand{\cpl}[3]{\jcite{Chem. Phys. Lett.}{#1}{#2}{#3}}
\newcommand{\cp}[3]{\jcite{Chem. Phys.}{#1}{#2}{#3}}
\newcommand{\pr}[3]{\jcite{Phys. Rev.}{#1}{#2}{#3}}
\newcommand{\jpc}[3]{\jcite{J. Phys. Chem.}{#1}{#2}{#3}}
\newcommand{\jpca}[3]{\jcite{J. Phys. Chem. A}{#1}{#2}{#3}}
\newcommand{\jpcA}[3]{\jcite{J. Phys. Chem. A}{#1}{#2}{#3}}
\newcommand{\jcc}[3]{\jcite{J. Comput. Chem.}{#1}{#2}{#3}}
\newcommand{\molphys}[3]{\jcite{Mol. Phys.}{#1}{#2}{#3}}
\newcommand{\physrev}[3]{\jcite{Phys. Rev.}{#1}{#2}{#3}}
\newcommand{\mph}[3]{\jcite{Mol. Phys.}{#1}{#2}{#3}}
\newcommand{\cpc}[3]{\jcite{Comput. Phys. Commun.}{#1}{#2}{#3}}
\newcommand{\jcsfii}[3]{\jcite{J. Chem. Soc. Faraday Trans. II}{#1}{#2}{#3}}
\newcommand{\jacs}[3]{\jcite{J. Am. Chem. Soc.}{#1}{#2}{#3}}
\newcommand{\ijqcs}[3]{\jcite{Int. J. Quantum Chem. Symp.}{#1}{#2}{#3}}
\newcommand{\ijqc}[3]{\jcite{Int. J. Quantum Chem.}{#1}{#2}{#3}}
\newcommand{\spa}[3]{\jcite{Spectrochim. Acta A}{#1}{#2}{#3}}
\newcommand{\tca}[3]{\jcite{Theor. Chem. Acc.}{#1}{#2}{#3}}
\newcommand{\tcaold}[3]{\jcite{Theor. Chim. Acta}{#1}{#2}{#3}}
\newcommand{\jpcrd}[3]{\jcite{J. Phys. Chem. Ref. Data}{#1}{#2}{#3}}
\newcommand{\APJ}[3]{\jcite{Astrophys. J.}{#1}{#2}{#3}}
\newcommand{\astast}[3]{\jcite{Astron. Astrophys.}{#1}{#2}{#3}}
\newcommand{\arpc}[3]{\jcite{Ann. Rev. Phys. Chem.}{#1}{#2}{#3}}


\draft
\title{A definitive heat of vaporization of silicon through benchmark
ab initio calculations on SiF$_4$}
\author{Jan M.L. Martin$^*$}
\address{Department of Organic Chemistry,
Kimmelman Building, Room 262,
Weizmann Institute of Science,
76100 Re\d{h}ovot, Israel. {\em Email:} \verb|comartin@wicc.weizmann.ac.il|
}
\author{Peter R. Taylor}
\address{San Diego Supercomputer Center and Department of
Chemistry and Biochemistry~MC0505, University of California, San Diego, 
San Diego, CA 92093-0505, USA. {\em Email:~\verb|taylor@sdsc.edu|}}
\date{Submitted to {\it J. Phys. Chem. A} February 1, 1999}
\maketitle
\begin{abstract}
In order to resolve a significant uncertainty in the heat of 
vaporization of silicon --- a fundamental parameter in gas-phase 
thermochemistry --- $\Delta H^\circ_{f,0}$[Si($g$)] has been 
determined from a thermochemical cycle involving the precisely known
experimental heats of formation of SiF$_4$($g$) and F($g$) and a 
benchmark calculation of the total atomization energy (TAE$_0$) of
SiF$_4$ using coupled-cluster methods. 
Basis sets up to $[8s7p6d4f2g1h]$ on Si and $[7s6p5d4f3g2h]$
on F have been employed, and extrapolations for residual basis set 
incompleteness applied. The contributions of inner-shell correlation
($-$0.08 kcal/mol),  
scalar relativistic 
effects ($-$1.88 kcal/mol), atomic spin-orbit splitting ($-$1.97 kcal/mol), 
and anharmonicity in the 
zero-point energy (+0.04 kcal/mol) 
have all been explicitly accounted for. Our benchmark
TAE$_0$=565.89$\pm$0.22 kcal/mol leads to 
$\Delta H^\circ_{f,0}$[Si(g)]=107.15$\pm$0.38 kcal/mol
($\Delta H^\circ_{f,298}$[Si(g)]=108.19$\pm$0.38 kcal/mol):
between the
JANAF/CODATA value of 106.5$\pm$1.9 kcal/mol
and the revised value proposed by Grev and Schaefer [\JCP{97}{8389}{1992}],
108.1$\pm$0.5 kcal/mol. The revision will be relevant for future 
computational studies on heats of formation of silicon compounds. 
Among standard computational 
thermochemistry methods, G2 and G3 theory exhibit large errors,
while CBS-Q performs relatively well and the very recent W1 theory
reproduces the present calibration result to 0.1 kcal/mol.
\end{abstract}

\section{Introduction}

For three of the first-and second-row elements, namely Be, B, and Si,
the tabulated
heats of formation of the atoms in the gas phase carry experimental 
uncertainties in excess of 1 kcal/mol. Aside from being propagated into
uncertainties for experimental gas-phase thermochemical data for 
compounds involving these elements, they adversely affect the accuracy
of any {\em directly computed} heat of formation --- be it {\em ab initio} or
semiempirical 
--- of any Be, B, or Si-containing compounds through the identity
\begin{eqnarray}
\Delta H^\circ_{f,T}(\hbox{X$_k$Y$_l$Z$_m$\ldots})
   &-& k \Delta H^\circ_{f,T}(\hbox{X})
   - l \Delta H^\circ_{f,T}(\hbox{Y})
   - m \Delta H^\circ_{f,T}(\hbox{Z}) - \ldots\nonumber\\
=   E_T(\hbox{X$_k$Y$_l$Z$_m$\ldots}) &+& RT (1-k-l-m-\ldots)
   - k E_T(\hbox{X}) - l E_T(\hbox{Y}) - m  E_T(\hbox{Z}) - \ldots
\end{eqnarray}
Particularly given the importance of boron and silicon compounds, this
is a rather unsatisfactory state of affairs.

Recently we succeeded\cite{bf3} in reducing the uncertainty for boron
by almost an order of magnitude (from 3 kcal/mol to 0.4 kcal/mol)
by means of a benchmark calculation of the total
atomization energy (TAE$_0$) of BF$_3$(g). By combining the latter with the
experimentally precisely known\cite{Cod89} 
heat of formation of BF$_3$, we were able
to indirectly obtain the vaporization enthalphy of boron to high
accuracy. It was thus shown that a 1977 experiment by Storms and 
Mueller\cite{Sto77}, which was considered an outlier by the leading
compilation of thermochemical tables\cite{Jan98}, was in fact the 
correct value.

The heat of formation of Si($g$) is given in the JANAF\cite{Jan98} as
well as the CODATA\cite{Cod89} tables as 106.5$\pm$1.9 kcal/mol.
Desai\cite{Desai}
reviewed the available data and recommended the JANAF/CODATA value, but with a
reduced uncertainty of $\pm$1.0 kcal/mol. Recently, Grev and
Schaefer (GS)\cite{Gre92} found that their ab initio calculated TAE[SiH$_4$], 
despite basis set incompleteness, was actually {\em larger} than the
value derived from the experimental heats of formation of Si($g$), H($g$),
and SiH$_4$($g$). They suggested that the
heat of vaporization of silicon be revised upwards
to $\Delta H^\circ_{f,0}$[Si($g$)]=108.07$\pm$0.50 kcal/mol, a suggestion
supported by Ochterski et al.\cite{Och95}. 

The calculations by GS neglected relativistic contributions,
which were very recently considered by Collins and Grev (CG)\cite{Col98}. Using
relativistic (Douglas-Kroll\cite{DK})
coupled-cluster methods, these authors found that the TAE
of SiH$_4$ contains a relativistic contribution of $-$0.67 kcal/mol. 
Combined with the earlier calculations of GS, this yields 
$\Delta H^\circ_{f,0}$[Si($g$)]=107.4$\pm$0.5 kcal/mol, within Desai's
reduced error bar. However, as discussed there\cite{Col98}, the experimental
data for silane, SiH$_4$, involve an ambiguity. The JANAF heat
of formation of silane, 10.5$\pm$0.5 kcal/mol is in fact the Gunn and
Green\cite{Gun61} measurement of 9.5 kcal/mol increased with a 
correction\cite{Ros52} of +1 kcal/mol for the phase transition
Si(amorphous)$\rightarrow$Si(cr), which was considered an 
artifact of the method of
preparation by Gunn and Green. If one were to accept their argument, 
the GS and CG calculations on SiH$_4$ would actually support the original
JANAF/CODATA $\Delta H^\circ_{f,0}$[Si($g$)]. 

No such ambiguities exist for tetrafluorosilane, SiF$_4$, for which 
a very accurate experimental heat of formation has been determined\cite{Joh86}
by direct combination of the pure elements in their respective standard states
in a fluorine bomb calorimeter. Johnson's\cite{Joh86} heat of formation at
298.15 K, $-$386.18$\pm$0.11 kcal/mol, is slightly higher in absolute value
and slightly more precise than the CODATA value of $-$386.0$\pm$0.2 kcal/mol,
itself based on an earlier experiment from the same laboratory\cite{Wis63}.

Clearly, if a benchmark quality (preferably $\pm$0.3 kcal/mol or better)
TAE[SiF$_4$(g)] could be calculated, then
an unambiguous redetermination of $\Delta H^\circ_{f,0}$[Si($g$)] would
be possible. Our previous study on BF$_3$ being at the limit of the then
available computational hardware, a similar study on SiF$_4$ --- which 
contains an additional heavy atom and eight additional valence electrons,
leading to an expected increase in CPU time  and memory requirements 
by a factor of about 3.7 (see below) --- could only be completed most
recently, and is reported in the present contribution.

\section{Methods}

Most electronic structure calculations reported here were carried out using
MOLPRO 97.3\cite{molpro} running on SGI Octane and SGI Origin 2000
minisupercomputers at the Weizmann Institute of Science. The very
largest calculation, a full-valence coupled-cluster calculation
involving 620 basis functions, was carried out on
the National Partnership for Advanced Computational Infrastructure CRAY~T90 
at the San Diego Supercomputer Center.

As in our previous study on BF$_3$, all electron correlation calculations
involved in determining the valence
and inner-shell correlation contributions to TAE were carried out using the
CCSD\cite{Pur82} and CCSD(T)\cite{Rag89,Wat93} coupled-cluster
methods. (For the energies of the constituent atoms, the 
definition of Ref.\cite{Wat93} for the open-shell CCSD(T) energy
was employed.) Both the very low ${\cal T}_1$ diagnostic\cite{Lee89} 
of 0.012, and inspection of the largest coupled-cluster amplitudes,
suggest a system essentially totally dominated by dynamical correlation.
From experience it is known\cite{Lee95} that CCSD(T) yields results 
very close to the exact (full configuration interaction)
basis set correlation energy under such circumstances.

Basis set limits for the SCF and valence correlation limits were
extrapolated (see below for details) from calculated results using
the (A)VTZ+2d1f, (A)VQZ+2d1f, and (A)V5Z+2d1f basis sets. For
silicon, those basis sets consist of the standard Dunning
correlation consistent\cite{Dun89,Woo93} cc-pVTZ, cc-pVQZ, and
cc-pV5Z basis sets augmented with two high-exponent $d$ and one high-exponent
$f$ functions with exponents obtained by progressively multiplying the highest 
exponent already present by a factor of 2.5. The addition of such `inner
shell polarization functions'\cite{so2} has been shown\cite{so2,Bau95,sio,h2sio}
to be essential for smooth basis set convergence in second-row compounds, 
particularly those containing highly polar bonds such as
SiF$_4$\cite{Gil98}.  (It should be
recalled that inner shell polarization is a pure SCF effect and bears little
relationship to inner shell correlation. In the present case of SiF$_4$, 
the contribution of the inner polarization functions to the SCF/(A)VTZ+2d1f
TAE was found to be no less than 9.81 kcal/mol.)  For fluorine, the basis
sets given correspond to Dunning (diffuse function)-augmented correlation consistent\cite{Ken92}
aug-cc-pVTZ, aug-cc-pVQZ, and aug-cc-pV5Z basis sets --- it was shown 
repeatedly (e.g.\cite{l4}) that the use of augmented basis
sets on highly electronegative elements such as F in polar compounds is
absolutely indispensable for accurate binding energies. The final basis sets
for SiF$_4$ involve 235, 396, and 620 basis functions, respectively, for
(A)VTZ+2d1f, (A)VQZ+2d1f, and (A)V5Z+2d1f.

The geometry of SiF$_4$ was optimized by repeated parabolic interpolation
at the
CCSD(T)/cc-pVQZ+1 level, where the suffix `+1' stands for the addition of
a tight $d$ function with an exponent\cite{sio} 
of 2.082 on Si. In previous work on
H$_2$SiO\cite{h2sio}, one of us found that this recovers essentially all
of the inner polarization effect on the molecular geometry. The 
bond length thus obtained, $r_e$[SiF$_4$]=1.56043 \AA, was 
used throughout this work. (For comparison, the experimental $r_0$=1.5598(2)
\AA\cite{McD92}; to our knowledge, no experimentally derive $r_e$ is 
available.)

The inner-shell correlation contribution was determined by comparing the
computed binding energies correlating all electrons except Si(1s), and
correlating only valence electrons, using the MTsmall basis set\cite{W1}.
The latter is a variant of the Martin-Taylor core correlation basis set\cite{hf,cc} in which the very tightest $p$, $d$, and $f$ functions were deleted 
at no significant loss in accuracy on the contributions to TAE. 

The scalar relativistic contributions were obtained as expectation
values of the first-order Darwin and mass-velocity operators\cite{Cow76,Mar83}
at the ACPF (averaged coupled-pair functional\cite{Gda88}) level using the
MTsmall basis set. All electrons were correlated in this calculation, and
it should be noted that the MTsmall basis set is completely uncontracted and
therefore flexible enough in the $s$ and $p$ functions for this purpose.
For the sake of illustration, this approach yields $-$0.67 kcal/mol for
SiH$_4$, identical to two decimal places with the more rigorous 
relativistic coupled-cluster value\cite{Col98}.

The contribution of atomic spin-orbit splitting derived from the experimental
atomic fine structures\cite{Moo63} of Si($^3P$) and F($^2S$) is
$-$1.968 kcal/mol. For comparison, we also carried out all-electron 
CASSCF/CI spin-orbit calculations\cite{Hes95} 
using the $spdf$ part of
a completely uncontracted aug-cc-pV5Z basis set, augmented with a single
tight p, three tight d, and two tight f functions in even-tempered series with
ratio 3.0. In this manner, 
we obtain a contribution of $-$1.940 kcal/mol. In short, to the
accuracy relevant for this purpose it is immaterial whether the computed or
the experimentally derived value is used.

The zero-point energy was obtained from the experimentally derived
harmonic frequencies and anharmonicity constants of McDowell 
et al.\cite{McD92}. This leads to a value of 8.029 kcal/mol, whereas
one would obtain 8.067 kcal/mol from one-half the sum of the harmonic
frequencies, $\sum_i{d_i\omega_i}/2$
and 7.975 from one-half the sum of the fundamentals,
$\sum_i{d_i\nu_i}/2$. The approximation $\sum_i{d_i(\omega_i+\nu_i)/4}$,
at 8.021 kcal/mol, yields essentially the exact result.

\section{Results and discussion}

All relevant data are given in Table 1.

As expected, the SCF contribution of TAE converges quite rapidly.
We have shown previously\cite{nato} that the SCF convergence behavior
is best described by a geometric extrapolation $A+B/C^n$ of the type
first proposed by Feller\cite{Fel92}, with extrapolation from the
TAE contributions to be preferred over extrapolation from the constituent
total energies. From the (A)VTZ+2d1f, (A)VQZ+2d1f,
and (A)V5Z+2d1f results, i.e. Feller(TQ5), we obtain a basis set limit
of 448.43 kcal/mol, 0.02 kcal/mol more than the SCF/(A)V5Z+2d1f result
itself. An extrapolation from the (A)VDZ+2d, (A)VTZ+2d1f, and (A)VQZ+2d1f
basis sets would have yielded 448.47 kcal/mol, an increment of 0.22
kcal/mol over the (A)VQZ+2d1f result.

Given the large number of valence electrons, connected triple excitations
account for a rather small part of the binding energy: 9.61 kcal/mol at the
CCSD(T)/(A)VQZ+2d1f level, compared to a CCSD valence correlation
contribution of 114.85 kcal/mol and an SCF contribution of 448.25 kcal/mol.
Since a CCSD(T)/(A)V5Z+2d1f calculation is beyond the limits particularly
of memory and available CPU time for this system, this suggests an 
approach in which only the CCSD valence correlation contribution be obtained
from the largest basis set, while the (T) contribution is obtained from
an extrapolation on smaller basis sets. Indeed, Martin and de Oliveira (MdO)
recently found
in a systematic study\cite{W1} on a wide variety of first-and second-row
molecules that this essentially does not affect 
the quality of the results, except when the (T) contribution is a dominant
component to the binding energy. Helgaker and coworkers\cite{Hel97} previously
noted the more rapid basis set convergence behavior of connected triple
excitations as compared with the CCSD correlation energy.

The CCSD/(A)V5Z+2d1f calculation required over 3GB of memory, some
120 GB of disk space, and 43 hours of real time (81 hours of CPU time)
running on 8 CPUs of the NPACI CRAY~T90.
(Close to 99\% parallellism was achieved in the CCSD code simply
by adapting it to use vendor-supplied parallel BLAS and LAPACK libraries.)
To our knowledge,
this is the largest coupled-cluster calculation ever carried out using a
conventional algorithm. 

We have considered two extrapolation formulas based on the asymptotic
behavior of pair correlation energies\cite{Sch63,Kut92}, namely the
3-point extrapolation $A+B/(l+1/2)^\alpha$ due to Martin, and the
2-point extrapolation $A+B/l^3$ formula due to Helgaker and 
coworkers\cite{Hal98}. (In both formulas, $l$ stands for the 
maximum angular momentum present in the basis set.) MdO found\cite{W1} 
that both formulas tend to predict the same basis set limit if extrapolated
from sufficiently large basis sets, but that the limits predicted by
the $A+B/l^3$ formula are much more stable with respect to reduction of the
sizes of the basis sets used in the extrapolation. This is at least in
part related to the fact that the three-point extrapolation involves, of
necessity, one value with an even smaller $l$ than the two-point 
extrapolation. 

As an illustration, let us consider the BF diatomic which was used to
refine the BF$_3$ result\cite{bf3}. From the three-point $A+B/(l+1/2)^\alpha$
extrapolation applied to AV$n$Z ($n$=3,4,5) valence correlation 
contributions to $D_e$, we obtain 38.35 kcal/mol, compared to 38.76 kcal/mol
for AV$n$Z ($n$=4,5,6). In contrast, a $A+B/l^3$ extrapolation applied to
AV$n$Z ($n$=Q,5) yields 38.78 kcal/mol, just like AV$n$Z ($n$=5,6) does;
application to AV$n$Z ($n$=T,Q) results yields an overestimate of 
39.08 kcal/mol. 

In the present case, the $A+B/l^3$ formula predicts a CCSD limit
contribution to TAE[SiF$_4$] of 119.28 kcal/mol from the (A)VQZ+2d1f and 
(A)V5Z+2d1f results, with the extrapolation accounting for 2.27 kcal/mol
of the final result. For comparison, extrapolation from two smaller
basis sets, (A)VTZ+2d1f and (A)VQZ+2d1f, yields 119.62 kcal/mol, while
the $A+B/(l+1/2)^\alpha$ formula
applied to all three values yields a much smaller value of 118.87 kcal/mol.  

The (T) contribution is computed as 9.11 and 9.61 kcal/mol, respectively,
in the (A)VTZ+2d1f and (A)VQZ+2d1f basis sets: 
assuming $A+B/l^3$ behavior, this extrapolates
to a limit of 9.98 kcal/mol. We thus finally find a basis set limit valence
correlation contribution of 129.26 kcal/mol.

As expected, the Si(2s,2p) and F(1s) 
inner-shell correlation energy is quite substantial
in absolute terms, accounting for some 28\% of the overall correlation
energy excluding the very deep Si(1s) core. As we have seen in the past
for second-row molecules,
however, the differential contribution to TAE nearly cancels, in this 
case being only +0.08 kcal/mol. This contribution is definitely dwarfed 
by that of scalar relativistic effects, which as we noted we compute to be
$-$1.88 kcal/mol. 

Combining all of the above with the atomic spin-orbit correction noted in
the Methods section, we finally obtain a ``bottom-of-the-well'' TAE$_e$
of 573.92 kcal/mol; combined with the experimentally derived ZPE,
we obtain TAE$_0$=565.89 kcal/mol. 

Combining this with the CODATA heats of formation of F$(g)$ and
SiF$_4$($g$), we finally obtain $\Delta H^\circ_{f,0}$[Si($g$)]=107.34
kcal/mol. Using the more recent $\Delta H^\circ_{f,0}$[SiF$_4$(g)] instead,
this value is reduced to 107.15 kcal/mol.

In order to make an assessment of the probable error in these values, we
should consider both the uncertainty in the calculated TAE$_0$ and the
propagated experimental uncertainties in $\Delta H^\circ_{f,0}$[SiF$_4$(g)]
and  $\Delta H^\circ_{f,0}$[F(g)]. 
Using exactly the same method as we have employed,
MdO obtained a mean absolute error of 0.22 kcal/mol for a wide variety
of first-and second-row molecules, which dropped as low as 0.16 kcal/mol
when some molecules with significant nondynamical correlation effects
were eliminated. Erring on the side of caution, we assign 0.22 kcal/mol 
as a standard deviation rather than an upper limit to the error.
Given uncertainties of 0.07 and 0.20 kcal/mol in the CODATA heats of 
formation for F$(g)$ and SiF$_4$(g), respectively, we obtain
107.34$\pm$0.41 kcal/mol for $\Delta H^\circ_{f,0}$[Si(g)]. Employing
the more recent Johnson\cite{Joh86} $\Delta H^\circ_{f}$[SiF$_4$(g)]
instead, which has a smaller uncertainty, we propose 
$\Delta H^\circ_{f,0}$[Si(g)]=107.15$\pm$0.38 kcal/mol as our final
estimate. (At 298.15 K, using the CODATA $H_{298}-H_0$ functions, this
corresponds to 108.19$\pm$0.38 kcal/mol.) 

Our final estimate is in fact within the reduced error limits of 
Desai\cite{Desai}, $\Delta H^\circ_{f,0}$[Si(g)]=106.5$\pm$1.0 kcal/mol. 
It agrees to within combined uncertainties with the GS value after 
applying CG's relativistic correction, 107.4$\pm$0.5 kcal/mol, which suggests
that the `spurious' Si(cr)$\rightarrow$Si(amorph) transition enthalpy 
discussed in the introduction may indeed have been a fair estimate.
In previous calculations\cite{Bau98,Bau97} 
on SiF$_4$ and SiCl$_4$, respectively, Bauschlicher and coworkers 
derived values of 107.5$\pm$2 and 107.8$\pm$2 kcal/mol, respectively,
in which the error bars are very conservative. In the context of a review
article\cite{nato} on high-accuracy theoretical thermochemistry, Martin
recently repeated the GS calculation on SiH$_4$ using techniques similar
to those employed here, and obtained a TAE$_0$[SiH$_4(g)$] consistent
with $\Delta H^\circ_{f,0}$[Si(g)]=107.55$\pm$0.5 kcal/mol if the 
Si(cr)$\rightarrow$Si(amorph) phase transition enthalpy was indeed included.
We conclude that all data support a slight increase in 
$\Delta H^\circ_{f,0}$[Si(g)] to the 107.15$\pm$0.38 kcal/mol value 
proposed in the present work.


As a final note, we consider the performance of some `standard' theoretical 
thermochemistry methods for this molecule, compared to our benchmark 
TAE$_e$=573.92$\pm$0.22 kcal/mol. As noted previously\cite{Pop97},
G2 theory\cite{g2} fails dismally, underestimating TAE$_0$ by 8.2 kcal/mol
even as both spin-orbit splitting and scalar relativistics were neglected,
which would together have increased the gap by a further 3.85 kcal/mol.
G3 theory\cite{g3}
represents a substantial improvement, being 2.2 kcal/mol below
our value including spin-orbit corrections: applying the scalar relativistic
correction to their value (or, equivalently, deleting it from our own 
calculation) would however increase that gap to a still substantial 4.1 
kcal/mol. Interestingly, both CBS-Q and CBS-QB3\cite{cbs} predict much
higher values, 576.0 and 577.0 kcal/mol, respectively. Neither value
includes spin-orbit or relativistic corrections: upon applying them,
we find that they underestimate our best result by only $-$1.8 and $-$0.8 
kcal/mol, respectively. Finally, the W1 theory very recently
proposed by Martin and de Oliveira\cite{W1} yields a value of 573.85 
kcal/mol, only 0.07 kcal/mol below the present calibration result. (W1
theory includes both scalar relativistic and spin-orbit corrections as
standard parts of the method.)

\section{Conclusions}

From an exhaustive ab initio calibration study on the SiF$_4$ 
molecule, we obtain a total atomization energy at 0~K of
565.89$\pm$0.22 kcal/mol. This value includes rather substantial
scalar relativistic ($-$1.88 kcal/mol) and atomic spin-orbit ($-$1.97 kcal/mol)
effects, as well as more minor effects of
inner-shell correlation ($-$0.08 kcal/mol)
and anharmonicity in the 
zero-point energy (+0.04 kcal/mol).
In combination with experimentally
very precisely known heats of formation of F($g$) and SiF$_{4}$($g$), 
we obtain
$\Delta H^\circ_{f,0}$[Si(g)]=107.15$\pm$0.38 kcal/mol
($\Delta H^\circ_{f,298}$[Si(g)]=108.19$\pm$0.38 kcal/mol).
This confirms the suggestion of  Grev and Schaefer\cite{Gre92} that
the rather uncertain JANAF/CODATA value of 106.5$\pm$1.9 kcal/mol
should be revised upward, albeit to about 1 kcal/mol lower than their
suggested 108.1$\pm$0.5 kcal/mol.
The revision will be relevant for future 
computational studies on heats of formation of silicon compounds. 
Among standard computational 
thermochemistry methods, G2 and G3 theory exhibit large errors,
while CBS-Q performs relatively well and the very recent W1 theory
reproduces the present calibration result to 0.1 kcal/mol.

\acknowledgments

JM is a Yigal Allon Fellow, the incumbent of the Helen and Milton
A. Kimmelman Career Development Chair (Weizmann Institute), and
an Honorary Research Associate (``Onderzoeksleider
in eremandaat'') of the
National Science Foundation of Belgium (NFWO/FNRS).
This research was supported by the Minerva Foundation, Munich,
Germany~(JM), by the National
Science Foundation (USA) through Cooperative Agreement DACI-9619020 and 
Grant No. CHE-9700627~(PRT), and by a grant of computer time from SDSC.
The authors thank Drs. C. W. Bauschlicher Jr. and T. J. Lee (NASA Ames
Research Center) as well as 
Drs. R.D. Johnson III, P.A.G. O'Hare, and particularly K. K. Irikura (NIST) 
for helpful discussions, and Victor Hazlewood for assistance with
running the largest CCSD calculation reported here.

\begin{table}
	\caption{Computed thermochemical properties for SiF$_{4}$ and Si in 
	the gas phase. All values are in kcal/mol}
	\squeezetable
	\begin{tabular}{lddd}
			\multicolumn{4}{l}{Components of TAE}\\
			\hline
			 & SCF & CCSD-SCF & CCSD(T)-CCSD  \\
			\hline
			(A)VDZ+2d & 429.45 & 100.39 & 6.03  \\
			(A)VTZ+2d1f & 446.41 & 108.31 & 9.11  \\
			(A)VQZ+2d1f & 448.25 & 114.85 & 9.61  \\
			(A)V5Z+2d1f & 448.41 & 117.01 & ---  \\
			Extrap.\{D,T,Q\} & 448.47 & 119.62 & 9.98  \\
			Extrap.\{T,Q,5\} & 448.43 & 119.28 & ---  \\
			\hline
			\multicolumn{4}{l}{Best estimates:}\\
			\hline
			valence correlation & 129.26\\
			inner-shell correlation & 0.08\\
			Darwin\&mass-velocity & $-$1.88\\
			Atomic fine structure & $-$1.97\\
			best TAE$_{e}$ & 573.92\\
			ZPVE & 8.03 \\
			best TAE$_{0}$ & 565.89 \\
			\hline
			\multicolumn{4}{l}{Derivation of revised $\Delta H^\circ_{f,0}$[Si(g)]}\\
			 & $\Delta H^\circ_{f,298}$ & $H_{298}-H_{0}$ & $\Delta H^\circ_{f,0}$  \\
			\hline
Si($cr$) \cite{Cod89}      & 0 & 0.769$\pm$0.002 &  0 \\
Si($g$) \cite{Cod89}       & 107.6$\pm$1.9 & 1.8045$\pm$0.0002 & 106.5$\pm$1.9  \\
SiF$_{4}$($g$) \cite{Cod89}& $-$386.0$\pm$0.2 & 3.67$\pm$0.01 & $-$384.7$\pm$0.2  \\
SiF$_{4}$($g$) \cite{Joh86}& $-$386.18$\pm$0.11 & --- & 
$-$384.86$\pm$0.13$^{a}$  \\
F($g$) \cite{Cod89}        & 18.97$\pm$0.07 & 1.5578$\pm$0.0002 & 18.47$\pm$0.07  \\
F$_{2}$($g$) \cite{Cod89}  & 0 & 2.1092$\pm$0.0002 & 0  \\
\hline
Si($g$) this work          & 108.19$\pm$0.28 & --- & 107.15$\pm$0.38$^{a}$\\
\end{tabular}

(a) CODATA values\cite{Cod89} for $H_{298}-H_{0}$ have been employed

\end{table}

\end{document}